\documentstyle[sprocl,epsfig]{article}

\def\be{\begin{equation}}
\def\ee{\end{equation}}
\def\bea{\begin{eqnarray}}
\def\eea{\end{eqnarray}}

\begin{document}

\begin{flushright}
June 18 1997\\
TAUP 2435/97
\end{flushright}

\title{ Lessons and Puzzles of DIS at low $x$ ( high energy )  }

\author{ Eugene Levin
${}^{\dagger\,*}$\footnotetext{${}^{\dagger}$ Invited talk given at
Conference
on
Perspective in
Hadronic Physics, Trieste, Italy, 12 - 16 May,1997.} 
\footnotetext{$^{*}$ Email: leving@ccsg.tau.ac.il .} }

\address{School of Physics and Astronomy\\
 Raymond and Beverly Sackler Faculty of Exact Science\\
 Tel Aviv University, Tel Aviv, 69978, ISRAEL\\
 and\\
 Theory Department, Petersburg Nuclear Physics Institute\\
 188350, Gatchina, St. Petersburg, RUSSIA}


\maketitle\abstracts{ This talk is a rather sceptical review of our
knowledge and understanding of deep inelastic scattering
at low $x$ ( high energy). We show that the well known success of the
DGLAP evolution equations in describing of experimental HERA data brought
more problems and puzzles than answers. We advocate that more systematic
theoretical and experimental investigations of the nonperturbative QCD
are needed to clarify physics of DIS at high energy.}
  
\section{Introduction}

In this talk we   will  answer two main questions:

$\bullet$ What have we  learned from  deep inelastic scattering ( DIS ) at
high energy ( low $x$ ) from HERA?

$\bullet$ What problems in DIS are still a challenge for QCD?

Trying to find these answers we present here a critical  review of our
knowledge and  understanding of DIS at high energy (low $x$). The motto of
this talk is:

{\em The well advocated success of the DGLAP evolution equation
\cite{DGLAP}  in
describing  HERA data in the region of low $x$ brought more problems and
puzzles than answers. To sort out  these problems, we need to know more
about  nonperturbative QCD at small distances.}

\section{Basics of QCD in DIS}
DIS occurs at small distances and this is the process most suitable  to
apply perturbative QCD (pQCD ).  The advantage  of DIS is the fact that we
have
three general approaches:  the renormalization group approach (RG), the
Wilson Operator Product Expansion ( WOPE \cite{WOPE}) and the
factorization theorem ( FT \cite{FACTOR} ), these provide  a deeper
insight to
the main properties of DIS, than any sophisticated calculation in pQCD.

{\bf RG:} Let us assume that we have integrated out all degrees of freedom
with
transverse momenta $k_t < \Lambda$ and obtained the effective Lagrangian:
\be \label{1}
L_{eff} \,\,=\,\,\sum_i C^i (\Lambda ) \,O_i\,\,,
\ee
where $\Lambda$ is the scale which
separates the small and large momenta and $O_i$ is arbitrary operator. The
physical idea of RG is very simple, namely,
{\em physical observables ( dimensionless ) do not depend on
scale $\Lambda$}. This  means that introducing a new scale $\Lambda'$, we
 obtain a new effective Lagrangian with $C^i (\Lambda')$. The RG says
that $C^i (\Lambda')\,=\,F(C^i(\Lambda))$ with known function $F$. This
powerful method leads to the Dokshitser- Gribov - Lipatov - Altarelli -
Parisi \cite{DGLAP}  ( DGLAP ) evolution equations that play the  role of
the
Coulomb law in DIS.

{\bf WOPE:} This is a usual  way to separate the power - like corrections
($ \propto (\frac{1}{Q^2})^n $) from the logarithmic ones ( $\propto (
\alpha_S \ln Q^2)^n$ ), where $Q$ is the scale of hardness in our process
( the virtuality of the photon in DIS ). The WOPE maintains  that any
structure
function (say $F_2$) can be written in the form:
\be \label{2}
F_2(x,Q^2)\,\,=\,\,F^{LT}_2( x, \ln
Q^2)\,\,+\,\,\frac{1}{Q^2}\,\,F^{HT}_2(x, \ln Q^2)\,\,+\,\,O (
\frac{1}{Q^4})\,\,,
\ee
where the cross section of the virtual photon is equal to $\sigma\,=\,
\frac{4 \pi^2 \alpha_{em}}{Q^2}\,F_2(x,Q^2)$. Both $F^{LT}_2$ and
$F^{HT}_2$ depend only on $\alpha_S\ln Q^2$ and their $\ln Q^2$ dependence 
can be calculated in pQCD using the DGLAP evolution.  The
expansion
of Eq.(2) is valid in any order of pQCD.

{\bf FT:}  Allows us to separate the nonperturbative contribution
(parton densities, $f(x,\Lambda^2)$ ) at
large distances ($r_t > \frac{1}{\Lambda}$),  from the perturbative ones (
coefficient functions, $C(x,\ln(Q^2/\Lambda^2))$. For any structure
function in Eq.(2) the FT gives:
\be \label{3}
F^{LT}_2(x,Q^2)\,\,=\,\,\int d x'
C(\frac{x}{x'},\ln(Q^2/\Lambda^2))\,\cdot\, f(x',\Lambda^2)\,\,.
\ee
Coefficient function $C$ can be calculated in pQCD while parton densities
are the nonperturbative input in all our perturbative calculations.
\section{Success (!?)  of the DGLAP evolution equations}
\subsection{ What is the situation?}
The HERA experiment\cite{HERA} shows that the deep inelastic structure
function $F_2(x,Q^2)$ has a steep behaviour in the small $x$ region
($10^{-2}\,> \,x\,>\,10^{-5}$), even for very small virtualities ($Q^2
\,\approx \,1 \,GeV^2 $). Indeed, considering $F_2\,\propto\,x^{ -
\lambda}$ at low $x$, the HERA data is consistent with a  $\lambda$ which
changes from 0.15
at $Q^2 = 0.85\, GeV^2$ to 0.4 at  $Q^2 = 20\, GeV^2$. This steep
behaviour is well described \cite{ROBERT} in framework of the DGLAP
evolution equations,
by all groups doing the global fit of the data (
GRV\cite{GRV},MRS\cite{MRS} and CTEQ \cite{CTEQ}). No other ingredients
such as the BFKL Pomeron \cite{BFKL} and/or the shadowing corrections
(SC)\cite{GLR}
are needed to describe the experimental data starting from sufficiently low
virtualities $Q^2 \,\approx \,1.5\,GeV^2$. We now attempt to understand
what compromise has been made to obtain   a good description of the
data and what has
been actually done.
\subsection{What has been done?}
  Let me recall a standard procedure
of solving of the DGLAP evolution equations.  The first step: we introduce 
moments of the structure function, namely,  $x G(x,Q^2)\,=\,\frac{1}{2 \pi
i}\int_C e^{-\omega \,\ln(1/x)} \,M(\omega, Q^2) \,d \omega$, where
contour
$C$ is located to the right of all singularities of moment $M(\omega,
Q^2)$. The second step: we find the solution to the DGLAP equation
for moment
\be \label{4}
\frac{d M(\omega, Q^2)}{d \ln Q^2}\,\,=\,\,\gamma(\omega)\,M(\omega,
Q^2)\,\,.
\ee
The solution is
\be \label{5}
M(\omega, Q^2)\,\,=\,\,M(\omega,Q^2_0)\,
\cdot\,e^{\gamma(\omega)\,\ln(Q^2/Q^2_0)}\,\,.
\ee
Here $M(\omega,Q^2_0)$ is the nonperturbative input which should be taken
from experimental data or from ``soft" phenomenology ( model).
The third step: we find the solution for the parton structure function
using the inverse transform, namely:
\be \label{6}
xG(x,Q^2) \,\,=\,\,\int_C \,\frac{d \omega}{2 \pi i}\,\,e^{\omega
\,\ln(1/x)\,+\,\gamma(\omega)\,\ln(Q^2/Q^2_0)}\,M(\omega,Q^2_0)\,\,.
\ee
Therefore, to find a solution of the DGLAP equation we need to know the
nonperturbative input $M(\omega,Q^2_0)$ and the anomalous dimension
$\gamma(\omega)$, which we can calculate in perturbative QCD. We summarize
below what has been calculated for the gluon anomalous dimension. We
present the result as a table in which each element  gives the order of
the
magnitude of the perturbative term that has been calculated.  We use
brackets (...) or [...] to indicate terms that have not yet  been
calculated.

\begin{center}
\begin{tabular}{c c c c c c c c c c }
{\bf$ \gamma^G(\omega)$}& & & & & & & & & \\
$\alpha_S$: & & & & &$ \frac{\alpha_S}{\omega}$ & $\alpha_S$ & $\alpha_S
\,\omega$ & $\alpha_S\,\omega^2 $ & ... \\
$ \alpha^2_S$:  & & & & &$\frac{\alpha^2_S}{\omega}$ & $\alpha^2_S$ &
$\alpha^2_S\,\omega$& $\alpha^2_S\,\omega^2$ & ...\\
$\alpha^3_S$: 
 & & & & $[\frac{\alpha^3_S}{\omega^2}]$ &
$(\frac{\alpha^3_S}{\omega})$
&$(\alpha^3_S)$ & $(\alpha^3_S \,\omega)$ & $(\alpha^3_S\,\omega^2)$ &
...\\
$\alpha^4_S$: & & $\frac{\alpha^4_S}{\omega^4}$ & $
[\frac{\alpha^4_S}{\omega^3}]$ & $(\frac{\alpha^4_S}{\omega^2})$
&$ (\frac{\alpha^4_S}{\omega})$ & $(\alpha^4_S)$ & $(\alpha^4_S \,\omega)$
&
$(\alpha^4_S \,\omega^2) $& ...\\
$\alpha^5_S$: & $\frac{\alpha^5_S}{\omega^5}$ &
$[\frac{\alpha^5_S}{\omega^4}]$ & $(\frac{\alpha^5_S}{\omega^3})$
&$(\frac{\alpha^5_S}{\omega^2})$ & $(\frac{\alpha^5_S}{\omega})$&
$(\alpha^5_S)$
& $(\alpha^5_S \,\omega )$ & $(\alpha^5_S \,\omega^2) $ &...\\
\end{tabular}
\end{center}
\centerline{}  
\centerline{{\bf Table}}
\centerline{}
 We can now  see what has been done in the global fits. The value of the
anomalous dimension have been calculated in $\alpha^2_S$ order ( two
first rows in our table) and the nonperturbative input has been taken in
the form  $M(\omega,Q^2_0)\,\propto \,\frac{1}{\omega - \omega_0}$ with $
\omega_0\,\approx\,$ 0.2 - 0.3.   This means that the structure function
at $Q^2 = Q^2_0$ increases as $ x^{- \omega_0}$ at $x \rightarrow 0$
\footnote{Strictly speaking this statement is correct for two global
fits: MRS and CTEQ. The GRV fit  has a  different initial condition which
we 
will discuss later.}.
 \subsection{What is the price that must be paid?}
 To  understand the main feature of the low $x$ behaviour of the deep
inelastic structure functions, it is instructive to consider the
asymptotics of
$xG(x,Q^2)$ using  Eq.(6). This asymptotic is determined by the saddle
point in $\omega = \omega_S$ and the equation for $\omega_S$ is
\be \label{7}
\ln ( 1/x)\,\,+\,\,\frac{\gamma( \omega_S )}{ d
\omega}\,|_{\omega = \omega_S}\,\ln(Q^2/Q^2_0)\,\,=\,\,O\,\,.
\ee
Substituting $\gamma( \omega )\,=\,\frac{N_c \alpha_S}{\pi \omega}$, one
obtains $\omega_S\,=\,\sqrt{\frac{N_c
\alpha_S}{\pi}\,\cdot\,\frac{\ln(Q^2/Q^2_0)}{\ln(1/x)}}$ and 
$\omega_S \,\rightarrow 0$  in the region of low $x$. This means
that $\omega = \omega_0$ contributes to the integral and, therefore, the
energy behaviour  of the DIS processes  at high energy  is determined by
the initial
partonic distributions.  For lower energies the saddle point contributes
and the value of $\omega_S$  ( see Ref. \cite{EKL} for the values of  
$\omega_S$ in HERA kinematic region)   gives us the typical  scale of
$\omega$ in the anomalous dimension as given in the Table. It turns out
that
in the 
HERA kinematic region  corrections of  order of
$(\frac{\alpha_S}{\omega})^n$ should be important ( see Ref. \cite{EKL}).

\subsection{The first lesson:}
We can describe the experimental data on the deep inelastic structure
function using only the DGLAP evolution equation without any new
ingredients, provide we  assume, that the initial parton distributions 
increases as $\frac{1}{x^{\omega_0}}$ and such behaviour has to be
understood.   The ordinary procedure  of taking into account leading and
next to leading contributions to the anomalous dimension cannot be
justified since  other corrections of the order of
$(\frac{\alpha_S}{\omega})^n$ with $n >2$ are large and should be
included. An attempt to explain the steep initial distribution was made in
the GRV parameterization which starts from extremely low $Q^2 \approx 0.25
GeV^2$. It was shown that the DGLAP evolution can generate a power - like
increase of the parton distribution at $Q^2 \approx 1\,GeV^2$. It is an
interesting idea, but the open question is why, we can start the DGLAP
evolution with so small virtualities where we have no reason to assume
that only leading twist contribution is essential. We  recall that
the DGLAP evolution equations only  describe the  leading twist term in
Eq.(2).
\section{A unified BFKL and DGLAP description of $F_2$ data}
\subsection{Low $x$ resummation and the BFKL equation}
The contributions of the order of   $(\frac{\alpha_S}{\omega})^n$  in
$\gamma^G(\omega)$  are  given by  the BFKL equation \cite{BFKL} and were
 calculated in Ref.\cite{JAR}. Namely,
\be \label{8}
\gamma^G_{BFKL}(\frac{N_c \alpha_S}{\pi\,\omega})\,=\,\sum^{\infty}_{n =
1}
\,a_{n} (\frac{N_c
\alpha_S}{\pi \,\omega})^n\,\,,
\ee
where $\gamma^G_{BFKL}(\frac{N_c \alpha_S}{\pi\,\omega})$ is given by
iterations
of \cite{BFKL}
\be \label{9}
1\,\,=\,\,\frac{N_c\alpha_S}{\pi\,\omega}\,\chi(\gamma^G_{BFKL})\,\,\,\,\,
where\,\,\,\,\,\chi(\gamma)\,=\,2\,\psi(1)
\,-\,\psi(\gamma)\,-\,\psi(
1 - \gamma)\,\,.
\ee

The BFKL anomalous dimension of Eq.(8) can be written in the form
\be \label{10}
\gamma^G_{BFKL}(\frac{N_c
\alpha_S}{\pi\,\omega})\,|_{\omega\,\rightarrow\,\omega_L}\,=\,\frac{1}{2}
\,+\,\sqrt{\frac{\omega_L - \omega}{\Delta}}\,\,,
\ee
where $\gamma^G_{BFKL}(\omega = \omega_L)\,=\,\frac{1}{2}$.  One
can therefore see that the anomalous dimension cannot exceed the value
$\gamma =
\frac{1}{2}$ or, in other word, the BFKL anomalous dimension should be
essential in the kinematic region where $\gamma$  is close to
$\frac{1}{2}$.
\subsection{Where is the BFKL Pomeron?}
It is easy to find the kinematic region where the BFKL contribution become
sizeable, since in the region of low $x$ the solution of the evolution
equation has a 
semiclassical form:
\be \label{11}
x G(x,Q^2)\,\,=\,\,C \,\cdot\, (\frac{1}{x})^{<\omega>}\,\cdot\,(
\frac{Q^2}{Q^2_0})^{<\gamma>}\,\,,
\ee
where $C$,$<\omega>$ and $<\gamma>$ are smooth functions of $\ln(1/x)$ and
$\ln(Q^2/Q^2_0)$. In Fig.1 line $<\gamma>\,=\,\frac{1}{2}$ is plotted,
using the GRV parameterization, namely, $<\gamma>\,=\,\frac{d
\ln(xG^{GRV}(x,Q^2))}{d \ln(Q^2/Q^2_0)}$. One can see, that HERA data has
penetrated the region where the BFKL dymanics should be visible. On the
other hand, as we have discussed, the data could be described without any 
 contamination from BFKL.
\begin{figure}
\centerline{\psfig{file=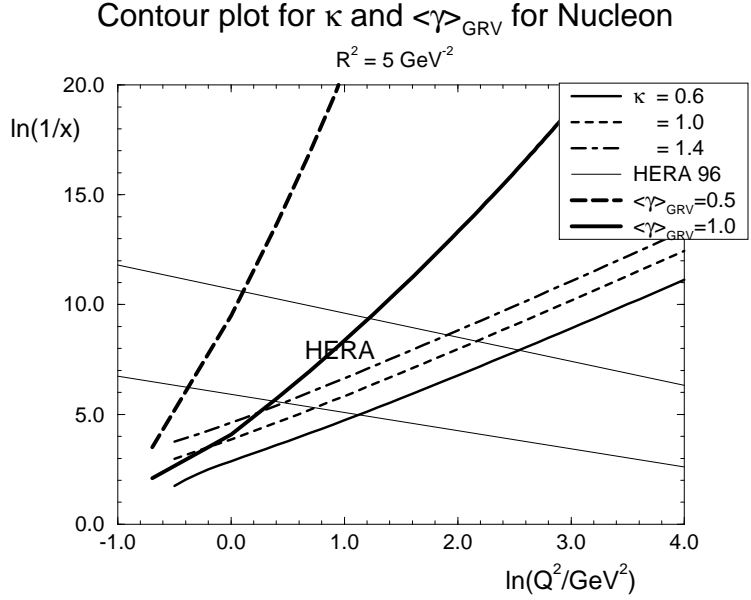,width=100mm}}
\caption{Lines with fixed $< \gamma>$ and $\kappa$ for the GRV
parametrization.}
\end{figure}
A possible answer has been given in two recent papers
\cite{THORNE}\cite{MKS}, in which an attempt has been made to describe the
HERA data using the following approach for
\be \label{12}
\gamma^G\,=\,\gamma^{\alpha_S}\,+
\,\gamma^{BFKL}(\frac{N_c\alpha_S}{\pi\,\omega}) \,\,.
\ee
 In other word, the first row  and all
terms of the order $ (\frac{N_c\alpha_S}{\pi\,\omega})^n$ in the table has
been included. Actually, such an approach was suggested  long ago in Ref.
\cite{GRV}, where the evolution equation  which corresponds to
Eq.(12) was written. The main result of these two papers  is  as
follows:
using Eq.(12) we can describe all data starting with the flat ($\omega_0\,
\rightarrow\,0$) initial distribution at $Q^2_0 \,\approx\,1\,GeV^2$. Such
initial conditions give a natural matching with the ``soft" processes.
The recent breakthrough in the calculation of the next order correction
to the BFKL equation \cite{FADIN} will lead to calculations in the nearest
future of all terms in
our table marked by [...]. It will allow us  to
calculate the next correction to Eq.(12) ($\Delta \gamma^G$), namely
\be \label{13}
\Delta \gamma^G\,=\,\gamma^{\alpha^2_S}\,+\,\gamma^{\alpha_S BFKL}\,=\,
\gamma^{\alpha^2_S}\,+\,\alpha_S\,\sum^{\infty}_{n =1}\,
(\frac{N_c\alpha_S}{\pi\,\omega})^n \,\,.
\ee
Using Eq.(13) we can evaluate the next order corrections and check how
well  our approach works.

\subsection{The second lesson:}
The widely  held  opinion that the DGLAP evolution equations describe the
experimental data is too biased. The equations that incorporate the BFKL
and DGLAP dynamics lead to better descriptions of the experimental data,
providing an excellent matching between the ``hard" and ``soft" processes
- flat initial parton distribution  at $Q^2 \,\approx \,1\,GeV^2$. 

\section{Higher twists}
Everything, that has been discussed, so far  is  related only  to the
leading twist
contribution to the deep inelastic structure function (first term in
Eq.(2)). The only way, to separate the leading twist from the
higher twist, is to consider sufficiently large value of $Q^2$,   where
the
higher twist contribution is expected to be small. However,  practical
estimates crucially depend on the $Q^2$ - dependence of $F^{HT}_2$. 
At first sight it appears that we do not know anything on  $F^{HT}_2$.
Actually, this is not
true. We know a lot about the next order  twist structure function:

1.  The physical meaning \cite{ELLISHT}, namely,the  high twist term is
closely  related to the correlation function of two partons in the parton
cascade.

2.The evolution equations \cite{BULI}. These equations are similar to  the
Fadeev equations for many body problem, namely, they reduce the
complicated parton interaction to the interaction of two partons  with
the same  kernel as in the  original  DGLAP equations.
 
3. The solution of the evolution equation  in the region of very
small $x$ ( high energy) \cite{HTLOWX}.

4. This solution suggests the following formula to fit the deep inelastic
structure function:
$$
F_2(x,Q^2)\,\,=\,\,F^{DGLAP}_2(x,Q^2)\,\,+\,\,\frac{m^2}{Q^2}c(x)\,
[\,F^{DGLAP}_2(x,Q^2)\,]^2\,\,,
$$ 

It should be stressed that this expression is quite different from that
one which experimentalists  assumed, namely, that $ F^{HT}_2$ has
the same $Q^2$ and $x$ dependence as $F^{LT}_2$.

However, nobody has tried to obtain  numerical estimates for the high
twist
contribution, as  the systematic computational approach to the
evolution equations for  $F^{HT}_2$  had not   been developed yet.   

This has  recently been  done \cite{BARTELSHT} and the result is  to some
extent
surprising. The extra power of $Q^2$ does not give the  feeling
that this
contribution should be negligibelly  small at least at $Q^2\,\approx\,10
GeV^2$. Indeed, if one  wants to try a simple parameterization for
$F^{HT}_2$
it is better to take $F^{HT}_2\,\propto\, ( F^{LT}_2)^2 )$ \cite{HTLOWX}  
\cite{BARTELSHT} and the $Q^2$ dependence due to anomalous dimension of
$F^{LT}_2$ compensates to a large extent for  the $1/Q^2$ suppression. The 
importance of this result is obvious, since in all solutions of the DGLAP
equations that there are on the market, it has been assumed that  the
higher
twist contribution is small, and can be neglected at $Q^2 = Q^2_0$, where
$Q^2_0$ is the initial virtually of the photon from which we start the
DGLAP evolution. Notice that in practice the value of  $Q^2_0$ is 
rather small ( about 4 $GeV^2$ ).

\subsection{ The  third lesson:}
The common believe that the higher  twist contributions are small at
$Q^2\,\approx\,4\,GeV^2$ does not look convincing. We have to study the
higher  twist contribution in detail  as to develop a systematic
 computational approach to the evolution equations for the higher twist
structure function.

\section{On the way to complete theory of DIS}
\subsection{Lattice calculations}
For the first time the lattice calculations in DIS  has achieved  an
accuracy
that we have to discuss them seriously. We would like to recall, that the
lattice experiment gives us  solid information on  the nonperturbative 
QCD contribution to DIS. In some sense this is the only way how we can
obtain a selfconsistent and reliable nonperturbative contribution.

The experimental errors in the lattice experiments are  still large  but,
nevertheless, it gives a convincing result that the initial quark
distributions at $Q^2 = 2.5\,GeV^2$ differs from experimental one ( see   
minireview \cite{LATTICE} at DIS'97). Fig.2 shows  the quark
distribution 
  predicted by lattice calculation ( see Ref. \cite{MAN}) and the same
distribution that the CTEQ collaboration used as an initial one. 
Such a
difference was expected since in the
lattice experiment the leading twist contribution has been calculated   
while  experimental data give the contribution of all twists at a
definite value of the virtuality $Q^2$. It is interesting to note that the
 leading twist quark distributions derived on the lattice turn out to be
closer to
one that was expected in the constituent quark model, where the mean
momentum
of
the quark   about $\frac{1}{3}$.
  
\begin{figure}
\centerline{\psfig{file=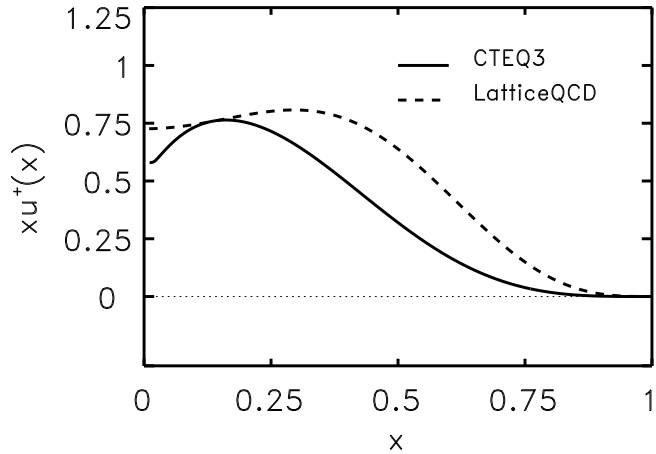,width=100mm}}
\caption{Reconstruction of u-quark distribution from lattice
calculation \protect\cite{MAN} ( dotted line) and the CTEQ
parametrization (solid line) at $\mu^2$ = 2.5 $GeV^2$.}
 \end{figure}

\subsection{ The fourth lesson:}
The lesson from the first nonperturbative calculation is very simple: our
usual method  of separation leading twist from the higher twist does not
work,
the higher twist contribution is still large at $Q^2 = Q^2_0 \,\approx\, 3
- 4 GeV^2$.
\subsection{A unique opportunity for a theoretical approach to DIS}
We would like to emphasize that the lattice calculations give an unique  
opportunity to develop a self consistent theoretical approach to  DIS. 
The following strategy is advocated .  First we  use the lattice parton
leading
twist densities as the initial conditions for the GLAP evolution
equations and  solve them. The difference between experimental initial
parton densities and the lattice one should be treated  as the
high twist contribution. For them we should use the high twist evolution
equations,, and  the theoretical  status of the higher  twist contribution
was
discussed. The above procedure will provide a theoretical  approach to
DIS and after it has been achieved  we will be able to discuss DIS on the
solid theoretical  basis.  However,  the experts in lattice  experiments
still  need   
to calculate the initial gluon density.
\section{Shadowing corrections: miracle or reality?}
The HERA data on deep inelastic structure function lead
to puzzling result. On one hand, they can be successfully described in
the framework of the DGLAP evolution equations without any new ingredients
like the BFKL equation and/or the shadowing corrections (SC),as have been 
discussed above. On the other
hand,  the parameter ($\kappa$) which gives the estimate for the
strength
of
the SC turns out to be large  ($\kappa\,\geq\,1$) in the HERA kinematic
region (see Fig.1). This parameter $\kappa$ was estimated in
Refs.\cite{GLR}
\cite{MUQI} and it is equal to
\be \label{14}
\kappa\,\,=\,\,xG(x,Q^2)\,\frac{\sigma(GG)}{\pi R^2}\,\,=\,\,x
G(x,Q^2)\,\frac{3 \pi \alpha_s}{Q^2\,R^2}\,\,.
\ee
To understand what is going on, we have to develop a theoretical approach 
in which we can treat the region of $\kappa \,\approx\,1$ in DIS. It
should
be stressed that  previous attempts to develop a theory for the SC  
\cite{GLR} only had
a guaranteed theoretical accuracy  for small $\kappa
\,\approx\,\alpha_s\,\ll\,1$. Two  such approaches were discussed
recently: in the first one
\cite{AGL}     pQCD was used at the edge of its validity
($\alpha_s \kappa\,\leq\,1$), while in the second
( see Refs.\cite{WEIGERT} ) the new approach was
developed  in the kinematic region of  high parton density QCD.

In Ref. \cite{AGL} a new evolution equation was derived which describes
 that each parton   in the parton cascade interacts with the target in
Glauber - Mueller approach \cite{MU90}.  The results are the following:   
(i)
$\kappa$ is the correct parameter that determines  the strength of the SC;
(ii) the SC   to the gluon structure function  are big   even in the  HERA
kinematic region, but nevertheless  the value of the shadowed gluon
structure function   is still within the experimental
errors or, another way of putting it, the difference between the shadowed
and
nonshadowed gluon structure functions does not exceed the difference   
between the gluon structure functions in the different parameterizations
such as the MRS,GRV and CTEQ ones; (iii) the SC to $F_2(x,Q^2)$
 in HERA kinematic region are so small  that they can be neglected;  (iv)
the
SC enter
the game before the BFKL equation and, therefore, the BFKL Pomeron cannot 
be seen in the deep inelastic structure function  since it is hidden
under substantial SC ( it is interesting to note that this result is seen 
just from Fig.1 where it is shown that the SC become important in the
kinematic region where $< \gamma >$ is still smaller than $\frac{1}{2}$);
and
  (v) in the
region of low $x$ the asymptotic behaviour of $xG(x,Q^2)$ is $xG(x,Q^2)\,
\rightarrow\, \frac{2\, R^2\, Q^2}{ \pi^2} \,\ln(1/x)\,\ln \ln (1/x)$. 
This
means that the gluon  density is  not saturated \cite{GLR}
unlike for the GLR equation.

The new approach has been developed  based on the idea of the
semiclassical gluon field  in the region of high parton density ( see
Refs.\cite{MCLV},\cite{WEIGERT} and \cite{KOVNER} ).
 The physical problem has been  pointed out a long ago ( see Ref.
\cite{GLR}and Ref.\cite{LALE} for updated review): at high energy (low
$x$ ) and /or for DIS with heavy nucleus we are dealing with the system of
partons so dense that conventional methods of pQCD does not work. However,
the typical distances are still small for DIS,    and this fact results in
weak correlations between partons, due to small the  coupling constant of 
QCD.

The revolutionary idea, suggested in Ref. \cite{MCLV} and developed in
Refs. \cite{WEIGERT} and \cite{KOVNER}, is: in the Bjorken frame for DIS 
we can replace the complex QCD interaction between parton in such a system
by the interaction of a parton $i$  with energy fraction $x_i$ with the
classical field created by all partons
with energy fraction $x$ bigger than $x_i$. Indeed, in leading log(1/x)
approximation of QCD  all parton with $x\,>\,x_i$ live for a  much longer 
time
than parton $i$, therefore, they create a gluon field which only depends
 on  
their density. Using this idea and Wilson renormalization group approach,
in  Refs.\cite{WEIGERT} and  \cite{KOVNER} ( see also Ref.\cite{KOVCH}
for elucidating  remarks), the effective Lagrangian was  obtained.

It has been demonstrated that this effective Lagrangian
correctly reproduces the DGLAP evolution equations \cite{WEIGERT} and even
the BFKL Pomeron \cite{KOVNER} in the limit of a sufficiently weak gluon
field. However, the main problem in  matching the  two approaches: one  
which we  discussed above in the beginning of this section  and this one,
is still an open
problem.
\section{Crazy ideas}
Unfortunately, only two ideas on high energy behaviour of DIS we could
call as crazy enough to be
interesting.

 The first one ( see Ref.\cite{BALL} )is a generalization of
the renormalization group approach to the longitudinal degrees of
freedom. The arguments are based on the $k_t$ -
factorization \cite{KTFACT} and on similarity between $\ln Q^2$ and $\ln
s$. The answer is, roughly speaking, the BFKL amplitude, but with
running
QCD coupling which depends on energy. Certainly, this answer does not
contradict  unitarity and, perhaps, even the experimental data, but, of
course, it is in strong contradiction to  the BFKL approach, that has
been discussed in section 4. None  of the experts
have  seen    a diagram  which survives at high energy and in which the
running
coupling constant depends on $s$, and  so it  makes the whole approach
rather
suspicious.  On the other hand, this approach is sure to stimulate   a
more detailed study of this problem.

The second idea is more physical and based on the successful Buchmuller -
Haidt parameterization of all data on photon - prpton interaction:
photoproduction and DIS \cite{BUHA}. Accordingly to Ref.\cite{BUHA}, the
best parameterization of the HERA data is not the solution of the DGLAP
evolution equations but a simple formula:
\be
F_2(x,Q^2)\,=\,m\,log\frac{(Q^2 +
Q^2_0)}{Q^2_0}\,\,log\frac{x_0}{x}\,\,,
\ee
with  m = 0.364; $x_0$ = 0.04; and $Q^2_0$ = 0.55 $GeV^2$.
Eq.(14) gives a correct limit for $F_2$ at $Q^2 \,\rightarrow \,0$ and
describes the experimental data on the photoproduction. Such a
parameterization leads to the gluon structure function \cite{BUHA}
$xG(x,Q^2)\,=\,3\,log\frac{x_0}{x}$, which energy ( $x$ ) dependence is
just the same as the solution to a new evolution equation which takes
into account the SC \cite{AGL}. The physical idea is really crazy and
sounds as follows: everything has happened at $x$ of the order of
$10^{-2}$, at such values of $x$ the parton densities have reached their
asymptotic values      and  in the region of $x \,\leq \,10^{-2}$ we see
only
the  asymptotic behaviour of the  parton densities at low $x$.

 \section{Resume}
I hope that I have  convinced you that the low $x$ behaviour of the deep
inelastic structure function is still an open problem and the success of
the DGLAP evolution equation caused  more problems than answers.

I believe, that I gave enough arguments to claim:

1. That corrections to the DGLAP anomalous dimension are not small in 
HERA kinematic region at low $x$   ;

2. That the higher  twist terms are not small at low $x$ even for
$Q^2\,\approx \,10 GeV^2$;

3. That the SC are not small in HERA kinematic region.

Therefore,
on one hand, we have to add the so called  BFKL anomalous dimension to
the DGLAP one in the region of small $x$ at HERA, but, on the other
hand, the BFKL evolution equation can not be detected due to  the
background of the strong SC.  The explanation of why such  different
pictures
can   describe the experimental data, is closely related to the lack of
 direct  measurement of the gluon structure function. Despites the
beautiful data on $F_2$ at HERA we only know the value of the gluon
structure
function within 100\% errors. 

The only way to decide between our theoretical approaches is to predict
the behaviour of the other processes at high energy, such as diffractive
production, inclusive cross section and correlations in DIS both for
nucleon and nuclear targets.  Howevere, we have to admit that the
 accuracy of our theoretical  calculations in all other processes, is not
as  good as for  the total cross section of DIS. 

The good news is that the success in lattice calculations allows  us 
 to a large extent to control  the nonperturbative contribution to DIS.
Based on the lattice calculation, we can develop a selfconsistent
theoretical approach ( see section 6.3) which enable us to clarify
some  of our problems in near future.

A special chapter of DIS is the ``hard" processes with a nuclear
target, which I did not touch upon here. They certainly will help us to
have  a better understanding of  both the higher twist contributions and
the SC, but
this is a subject for a separate talk.

\section*{Acknowledgments}
I am very grateful to my colleques at HEP department in the Tel Aviv
university and at DESY theory group for hot and provocative discussions
on the subject. Special thanks to Asher Gotsman for his comments on the
manuscript.


\section*{References}

\begin{thebibliography}{99}
\bibitem{WOPE}
K. Wilson: {\it Phys. Rev.} {\bf 179} (1969) 1444.
\bibitem{FACTOR}
 J.C. Collins, D.E. Soper and G. Sterman:{\it Nucl.Phys.}
{\bf B308} (1988) 833.
\bibitem{DGLAP}
V.N. Gribov and L.N. Lipatov:{\it Sov. J. Nucl. Phys.} {\bf 15} (1972)
438; L.N. Lipatov: {\it Yad. Fiz.} {\bf 20} (1974) 181; G. Altarelli and
G. Parisi:{\it Nucl. Phys.} {\bf B126} (1977) 298; Yu.L. Dokshitser:{\it 
Sov. Phys. JETP} {\bf 46} (1977) 641.
\bibitem{HERA}
H1 collaboration: S. Aid {\it et al.}, {\it Nucl. Phys.} {\bf B470}
(1996) 3;\\
ZEUS collaboration: M. Derrick {\it et al.},{\it Z.Phys.} {\bf C69}
(1996) 607, {\bf C72 } (1996) 399.
\bibitem{ROBERT}
R.G.Roberts:{\it Plenary talk at DIS'97}, Chicago, April 1997, 
RAL-TR-97-024,hep-ph/9706269.
\bibitem{GRV}
M. Gluck, E. Reya and A. Vogt: {\it Z.Phys.} {\bf C67}(1995)433.
\bibitem{MRS}
A.D. Martin, R.G. Roberts and W.J. Stirling: {\it
Phys.Lett.}{B306}(1993)145.
\bibitem{CTEQ}
 CTEQ Collaboration, H.L.Lai et al.: {\it Phys.Rev.}{\bf
D51}(1995) 4763.
\bibitem{BFKL}
E.A. Kuraev,  L.N. Lipatov and V.S. Fadin: {\it Sov. Phys. JETP} {\bf
45}(1977) 199 ;  \,\,Ya.Ya. Balitskii and L.V. Lipatov: {\it Sov.J.
Nucl. Phys.} {\bf 28} (1978) 822;\,\,L.N. Lipatov: {\it Sov. Phys. JETP}
{\bf 63} (1986) 904.
\bibitem{GLR}
L.V. Gribov, E.M. Levin and M.G. Ryskin: {\it Phys. Rep.} {\bf 100} 
 (1983) 1.
\bibitem{EKL}
R.K. Ellis, Z. Kunst and E. M. Levin: {\it Nucl. Phys.}{\bf
B420}(1994) 517.
\bibitem{JAR}
T. Jaroszewicz:{\it Phys. Lett.}{\bf B116}(1982)291.
\bibitem{THORNE}
R.S.Thorne: hep - ph /9701241 and references therein.
\bibitem{MKS}
 J. Kwiecinski, A.D. Martin and A.M.
Stasto: hep - ph /9703445.
\bibitem{FADIN}
V.Fadin: { \it Talk at DIS'97},
Chicago, April 1997.
\bibitem{ELLISHT}
R.K.Ellis,W. Furmanski and R. Petronzio: {\it Nucl.Phys.} {\bf B207}
(1982)1, {\bf B212} (1983) 29.
\bibitem{BULI}
A.P. Bukhvostov,G.V. Frolov, L.N. Lipatov and E.A. Kuraev: {\it Nucl.
Phys.} {\bf B258} (1985) 601.
\bibitem{HTLOWX}
J. Bartels: {\it Phys. Lett.} {\bf B298} (1993) 204, {\it Z. Phys.} {\bf
C60} (1993) 471; E.M. Levin, M.G. Ryskin and A.G. Shuvaev: {\it Nucl.
Phys.} {\bf B387} (1992) 589.
\bibitem{BARTELSHT}
J. Bartels: { \it Talk at DIS'97},
Chicago, April 1997.
\bibitem{LATTICE}
G. Schierholz: minireview, { \it Talk at DIS'97},
Chicago, April 1997.
\bibitem{MAN}
T.Weigl and L. Mankiewicz: {\it Phys. Lett.} {\bf B339} (1996) 334.
\bibitem{MUQI}  
A.H. Mueller and J. Qiu: {\it Nucl. Phys.}  {\bf B268} (1986) 427. 
\bibitem{AGL}
A.L. Ayala, M.B. Gay Ducati and E.M. Levin: TAUP 2432 - 97, hep -
ph/9706347; {\it Nucl. Phys.}{\bf B493}(1997) 305.
\bibitem{WEIGERT}
 J. Jalilian-Marian, A. Kovner, L. McLerran  and H. Weigert:
hep - ph /9606337; H. Weigert: {\it Talk at DIS'97},
Chicago, April 1997.
\bibitem{MU90}
A.H.Mueller: {\it Nucl. Phys.} {\bf B335} (1990) 115.
\bibitem{MCLV}
L. McLerran and R. Venugopalan: {\it Phys. Rev.} {\bf D49} (1994)
2233,3352, {\bf D50} (1994) 2225, {\bf D53} (1996) 458.
\bibitem{KOVNER}
 J. Jalilian-Marian, A. Kovner, A. Leonidov and H. Weigert: hep -
ph/9701284; A. Kovner:{\it Talk at DIS'97},
Chicago, April 1997.
\bibitem{LALE}
E. Laenen and E. Levin: {\it Ann. Rev. Nucl. Part.} {\bf 44} (1994) 199.
\bibitem{KOVCH}
Yu.V. Kovchegov: {\it Phys. Rev.} {\bf D54} (1996) 5463, hep -
ph/9701229; Yu.V. Kovchegov, A.H. Mueller and S. Wallon: hep -
ph/9704369.
\bibitem{BALL}
R.D. Ball and S. Forte:hep-ph/9703417; R.D. Ball:{ \it Talk at DIS'97},
Chicago, April 1997.
\bibitem{KTFACT}
S.Catani,M. Ciafaloni and F. Hautmann: {\it Phys. Lett.} {\bf B242}
(1990) 97, {\it Nucl. Phys.} {\bf B366} (1991) 135; J.C. Collins and
R.K.
Ellis: {\it Nucl. Phys.} {\bf B360} (1991) 3; E.M. Levin,M.G.
Ryskin,Yu.M.Shabelsky and A.G. Shuvaev: {\it Sov.J.Nucl.Phys.} {\bf 53}
(1991) 657.
\bibitem{BUHA}
W. Buchmueller and D. Haidt: DESY 96-061, D.Haidt:{ \it Talk at DIS'97},
Chicago, April 1997.
\end{thebibliography}
\end{document}